\documentclass[showpacs,aps,preprintnumbers,prb,amsmath,amssymb,twocolumn,floatfix]{revtex4-1}
\usepackage{graphicx}
\DeclareGraphicsExtensions{.pdf,.eps,.png,.jpg,.mps}
\begin{document}

\title{Superconductivity of La$_3$Co$_4$Sn$_{13}$ and La$_3$Rh$_4$Sn$_{13}$: A comparative study}

\author{A.~\'{S}lebarski$^{\star}$, M.~Fija\l kowski$^{\star}$, M.~M.~Ma\'ska$^{\star}$, M.~Mierzejewski$^{\star}$, B.~D.~White$^{\dag}$,  and M.~B.~Maple$^{\dag}$}
\affiliation{
$^{\star}$Institute of Physics,
University of Silesia, 40-007 Katowice, Poland\\
$^{\dag}$Department of Physics, University of California, San Diego, La Jolla,
California 92093, USA
}
\begin{abstract}
We report the electric transport and thermodynamic properties of the skutterudite-related La$_3$Co$_4$Sn$_{13}$ and La$_3$Rh$_4$Sn$_{13}$ superconductors.  Applying an external pressure to La$_3$Rh$_4$Sn$_{13}$, the resistive superconducting critical temperature $T_c$ decreases, while the critical temperature of La$_3$Co$_4$Sn$_{13}$ is enhanced with increasing pressure.  The positive pressure coefficient $dT_c/dP$ correlates with a subtle structural transition in La$_3$Co$_4$Sn$_{13}$ and is discussed in the context of lattice instabilities.  Specific heat data show that both compounds are typical BCS superconductors.  However, La$_3$Rh$_4$Sn$_{13}$ also exhibits a second superconducting phase at higher temperatures, which is characteristic of inhomogeneous superconductors.  We calculate the specific heat for an inhomogeneous superconducting phase, which agrees well with  experimental $C(T)$ data for La$_3$Rh$_4$Sn$_{13}$.  We also found that an applied pressure reduces this
  second superconducting phase.
\end{abstract}


\maketitle

\section{Introduction}

Recent systematic research on filled--cage compounds focuses on their thermoelectric properties due to low phonon thermal conductivities resulting from {\it rattling} of the atoms inside the cage.  In the case of the Ce--based filled--cage Kondo systems, thermoelectricity is also strongly enhanced at low temperatures as a result of sharp features in the electronic density of states at the Fermi energy.  Both effects are also expected in the series of skutterudite--related $R_3M_4$Sn$_{13}$ compounds, first reported by Remeika,{\it et al.} \cite{Remeika80} where $R$ is a rare--earth element and $M$ is a transition metal.  The discovery of superconductivity in La$_3M_4$Sn$_{13}$\cite{Israel05,Hodeau82,Kase2011} has attracted considerable attention and provided an avenue by which to better understand the relationship between superconductivity and magnetism in the presence of strong electron correlations.  The quasi-skutterudite superconducting compound Ca$_3$Ir$_4$Sn$_{13}$ is a good
  example of a correlated electron system with a superlattice quantum critical point (QCP), reported to emerge under chemical or physical pressure.\cite{Klintberg12}

La$_3M_4$Sn$_{13}$ compounds where $M$ = Co, Rh are characterized as BCS superconductors.\cite{Kase2011} In this work, we present a comprehensive thermodynamic and high-pressure electrical resistivity study on La$_3$Co$_4$Sn$_{13}$ and La$_3$Rh$_4$Sn$_{13}$.  We show evidence of nanoscale inhomogeneity as a bulk property of La$_3$Rh$_4$Sn$_{13}$ in the sense that the samples exhibit electronic disorder over length scales similar to the coherence length which cannot be removed by any standard annealing procedure.  Such a substantial nanoscale electronic inhomogeneity is characteristic of the bulk Bi$_2$Sr$_2$CaCu$_2$O$_{8+x}$ high-$T_c$ materials.

\section{Experimental details}

Polycrystalline La$_3$Co$_4$Sn$_{13}$ and La$_3$Rh$_4$Sn$_{13}$ samples have been prepared by arc melting the constituent elements on a water--cooled copper hearth in a high-purity argon atmosphere with an Al getter.  The samples were remelted several times to promote homogeneity and annealed at 870 $^{\circ}$C for 12 days.  Almost no mass loss ($\leq0.02$\%) occurred during the melting and annealing process.  All samples were carefully examined by x-ray diffraction analysis and found to be single phase with cubic structure (space group $Pm\bar{3}n$).\cite{Slebarski13}

Electrical resistivity $\rho$ was investigated by a conventional four-point ac technique using a Quantum Design physical properties measurement system (PPMS).  Electrical contacts were made with 50 $\mu$m-gold wire attached to the samples by spot welding.  Electrical resistivity measurements under pressure were performed in a beryllium-copper, piston-cylinder clamped cell.  A 1:1 mixture of $n$-pentane and isoamyl alcohol in a teflon capsule served as the pressure transmitting medium to ensure hydrostatic conditions during pressurization at room temperature.  The local pressure in the sample chamber was inferred from the inductively determined, pressure-dependent superconducting critical temperature of high-purity Sn.\cite{Smith69}

Specific heat $C$ was measured in the temperature range $0.4-300$ K and in external magnetic fields up to 9 T using a Quantum Design PPMS platform.  Specific heat $C(T)$ measurements were carried out on plate-like specimens with masses of about $10-15$ mg utilizing a thermal-relaxation method.  The dc magnetization $M$ and magnetic susceptibility $\chi$ results were obtained using a commercial superconducting quantum interference device magnetometer from 1.8 K to 300 K in magnetic fields up to 7 T.

\section{Results and discussion: a comparative study}

\subsection{Magnetic properties near the critical temperature $T_c$}

\begin{figure}[h!]
\includegraphics[width=0.48\textwidth]{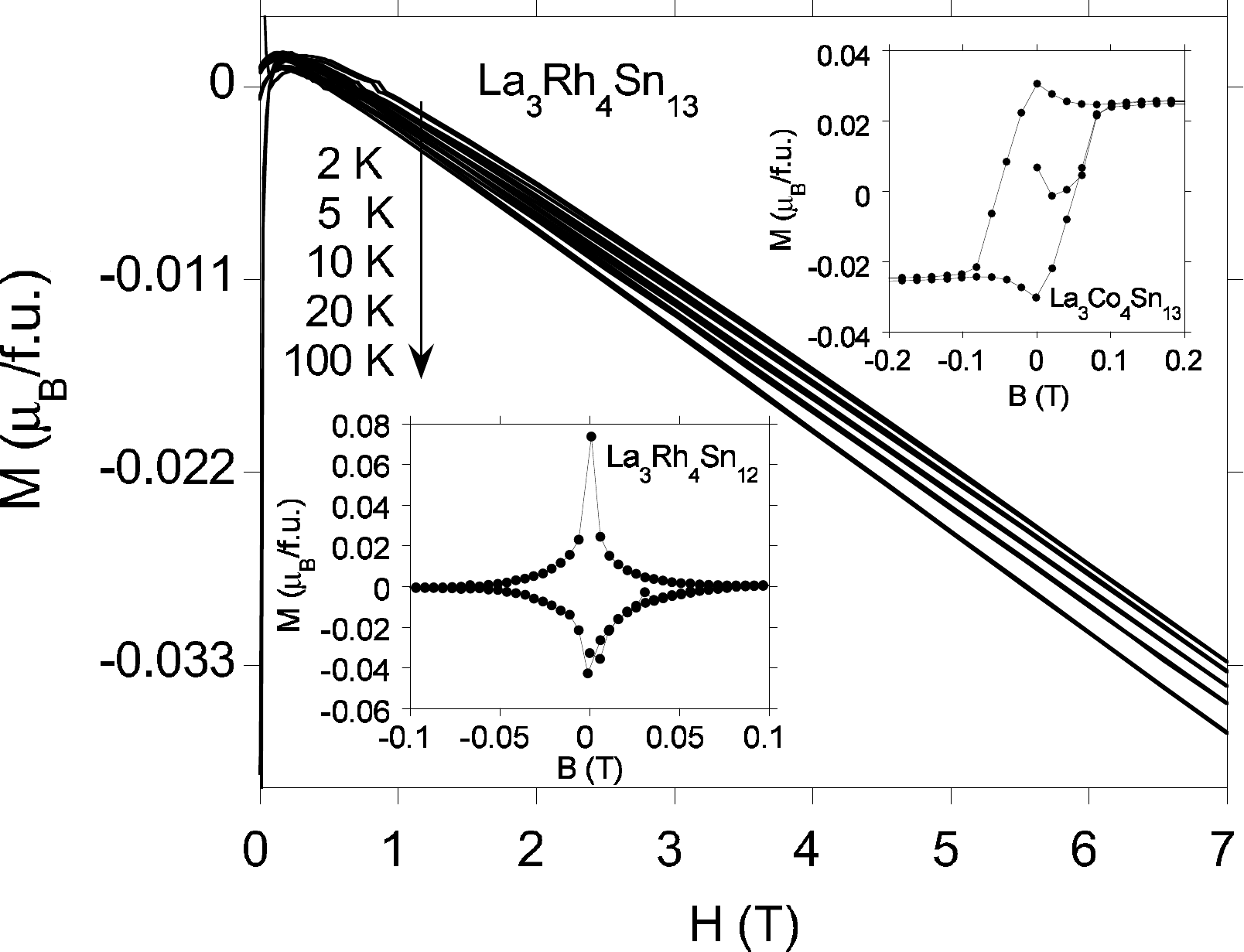}
\caption{\label{fig:Fig1}
Magnetization $M$ per formula unit vs magnetic field $B$ measured at different temperatures for La$_{3}$Rh$_{4}$Sn$_{13}$.  A very similar $M(B)$ dependence is observed for La$_{3}$Co$_{4}$Sn$_{13}$; therefore, the data are not presented here.  The insets display a symmetric hysteresis loop at $T = 1.8$ K in the superconducting state of La$_{3}$Rh$_{4}$Sn$_{13}$ and La$_{3}$Co$_{4}$Sn$_{13}$.}
\end{figure}

Figure~\ref{fig:Fig1} shows the magnetization $M$ vs $B$ isotherms for La$_{3}$Rh$_{4}$Sn$_{13}$.  A very similar $M(B)$ dependence was obtained for La$_{3}$Co$_{4}$Sn$_{13}$; therefore, the data are not displayed here.  The $M(B)$ isotherms are characteristic of a diamagnetic material with a small paramagnetic (or spin fluctuation\cite{Slebarski2012}) component.  The figure also displays a symmetric hysteresis loop at $T = 1.9$ K for La$_{3}$Rh$_{4}$Sn$_{13}$ and La$_{3}$Co$_{4}$Sn$_{13}$, characteristic of irreversible superconductivity.

\begin{figure}[h!]
\includegraphics[width=0.48\textwidth]{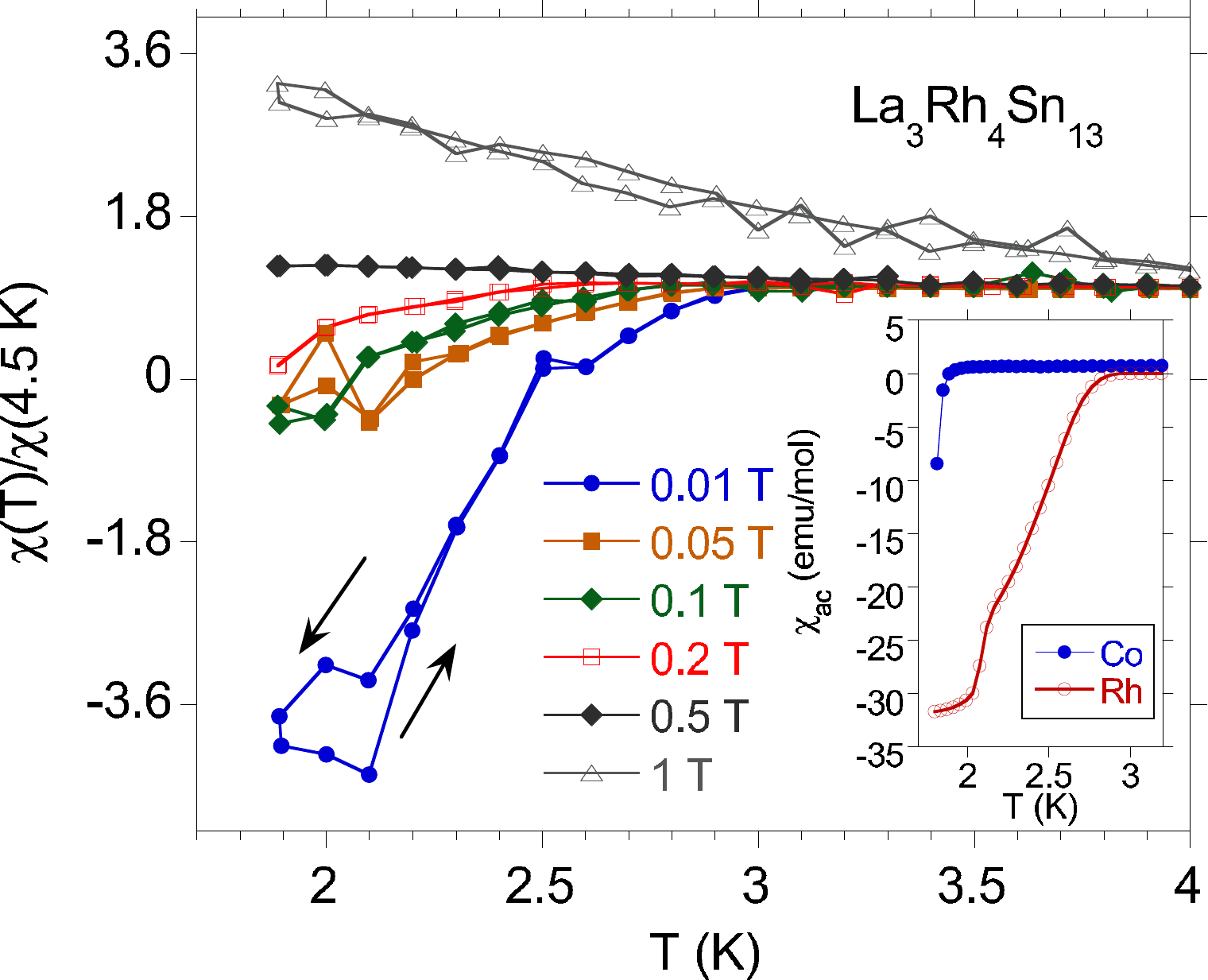}
\caption{\label{fig:Fig2}
(Color online) Magnetic susceptibility (dc) for La$_{3}$Rh$_{4}$Sn$_{13}$ at different magnetic fields measured with decreasing and increasing temperature.  In the superconducting state, there is a hysteresis loop which is strongly reduced by magnetic field.  The inset shows ac magnetic susceptibility measured in an applied ac magnetic field with amplitude of 1 Gs.  $\chi_{ac}$ of the Co-sample clearly shows a homogeneous superconducting state below $T_c$, while for La$_{3}$Rh$_{4}$Sn$_{13}$, it suggests an inhomogeneous superconducting phase between $T_c^{\star}$ and $T_c$.}
\end{figure}

Figure~\ref{fig:Fig2} displays the dc magnetic susceptibility obtained at different magnetic fields when the temperature is decreasing and then increasing, with hysteresis loops below $T_c$ for the applied magnetic fields $B\leq 0.1$ T.  Under applied magnetic fields larger than 0.5 T, the diamagnetism of the superconducting state is suppressed.
The ac magnetic susceptibility, displayed in the inset to Fig.~\ref{fig:Fig2}, clearly exhibits a homogeneous superconducting phase for La$_{3}$Co$_{4}$Sn$_{13}$, while for La$_{3}$Rh$_{4}$Sn$_{13}$, it shows evidence of two superconducting phases: an inhomogeneous superconducting state below $T^{\star}_{c} = 2.85$ K and a superconducting phase below $T_{c} = 2.1$ K with maximum diamagnetic $\chi_{ac}$ value. The $T^{\star}_{c}$ ``high-temperature'' phase will be discussed below.

\subsection{Electrical resistivity and specific heat under applied magnetic field}

\begin{figure}[h!]
\includegraphics[width=0.48\textwidth]{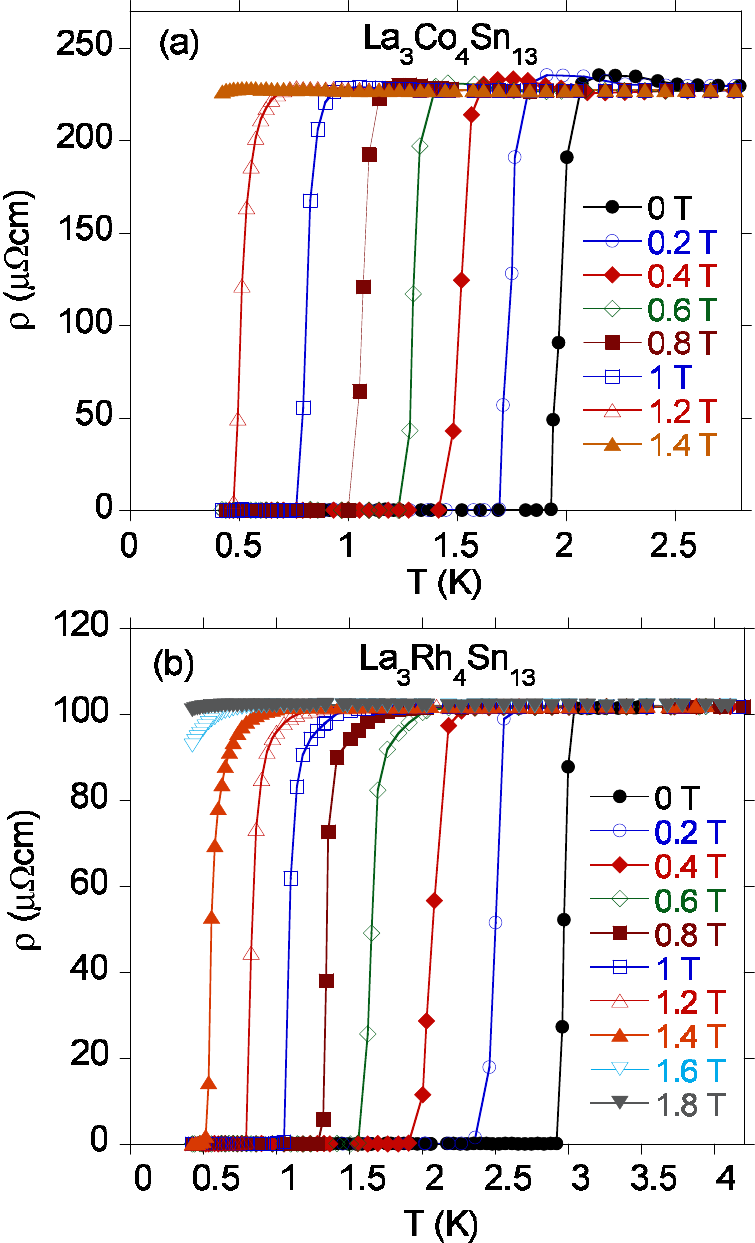}
\caption{\label{fig:Fig3}
(Color online) Temperature-dependent electrical resistivity $\rho$ of La$_{3}$Co$_{4}$Sn$_{13}$ (a) and La$_{3}$Rh$_{4}$Sn$_{13}$ (b) at various externally applied magnetic fields.  For clarity, the data are presented with a field increment 0.2 T.
}
\end{figure}

\begin{figure}[h!]
\includegraphics[width=0.48\textwidth]{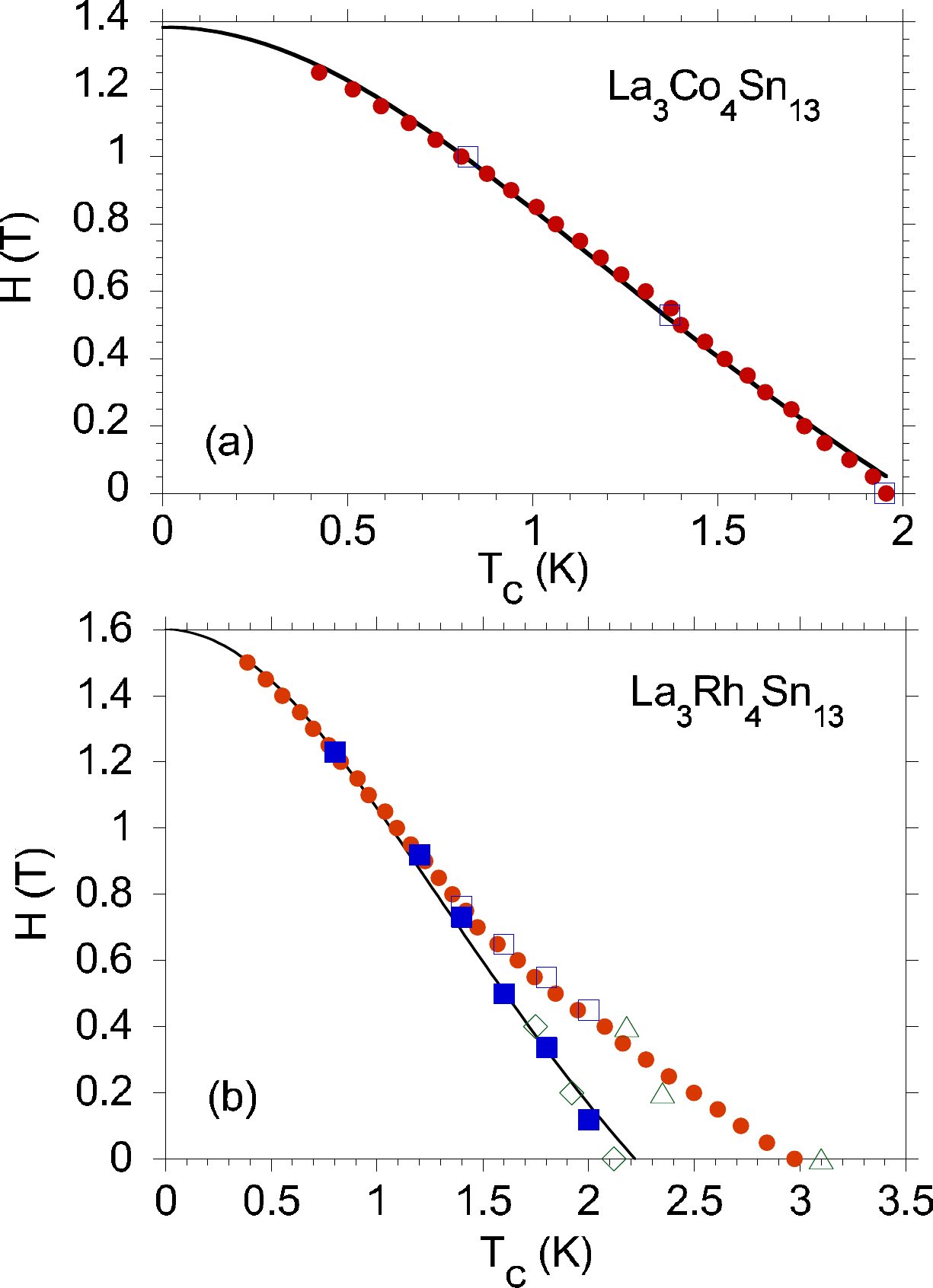}
\caption{\label{fig:Fig4} (Color online) Temperature dependence of the upper critical fields $H_{c2}$ in the $H-T$ phase diagram for La$_{3}$Co$_{4}$Sn$_{13}$ (a) and La$_{3}$Rh$_{4}$Sn$_{13}$ ($b$).  The solid line represents a fit using the Ginzburg-Landau model of $H_{c2}(T)$.  In panel ($b$), $T_c$ values characterized by red filled circles are obtained from electrical resistivity data under $H$, and defined as the temperature where $\rho$ drops to 50\% of its normal-state value.  The diamond and triangle data points represent $T_c$ and $T_c^{\star}$ values, respectively, obtained in a plot of $C/T$ vs $T$ in Fig.~\ref{fig:Fig5} on the line $H = {\rm const}$.  For the $T = {\rm const}$ line, the blue filled and unfilled squares represent the temperature of the break points in the plot of $C(T = {\rm const})$ vs $H$ (as shown in Fig.~\ref{fig:Fig6}).
}
\end{figure}

Figure~\ref{fig:Fig3} displays the temperature dependence of the electrical resistivity under applied magnetic fields for La$_{3}$Co$_{4}$Sn$_{13}$ and La$_{3}$Rh$_{4}$Sn$_{13}$.  We define $T_c$ as the temperature at 50\% of the normal-state resistivity value.  In Fig.~\ref{fig:Fig4}, we present the $H-T$ phase diagrams, where $T_c$s are obtained from electrical resistivity under several magnetic fields.
The Ginzburg-Landau (GL) theory approximates the $H-T$ diagram for La$_{3}$Co$_{4}$Sn$_{13}$ well.  In the upper panel, the best fit of the equation $H_{c2}(T)=H_{c2}(0)\frac{1-t^2}{1+t^2}$, where $t=T/T_c$ gives a value for the upper critical field $H_{c2}(0)=1.38$ T.  The upper critical field $H_{c2}$ can be used to estimate the coherence length.  Within the weak-coupling theory\cite{Schmidt77}
$\mu_0 H_{c2}(0)=\Phi_0/2\pi \xi_0^2$, so that the coherence length is estimated to be $\xi_0=16$ nm (the flux quantum $\Phi_0=h/2e=2.068 \times 10^{-15}$ T m$^2$).  However, in case of La$_{3}$Rh$_{4}$Sn$_{13}$ [Fig.~\ref{fig:Fig4}($b$)], its upper critical field curve evidently deviates from the GL theory in the fields $H > 1$ T.  This behavior is discussed below.

\begin{figure}[h!]
\includegraphics[width=0.48\textwidth]{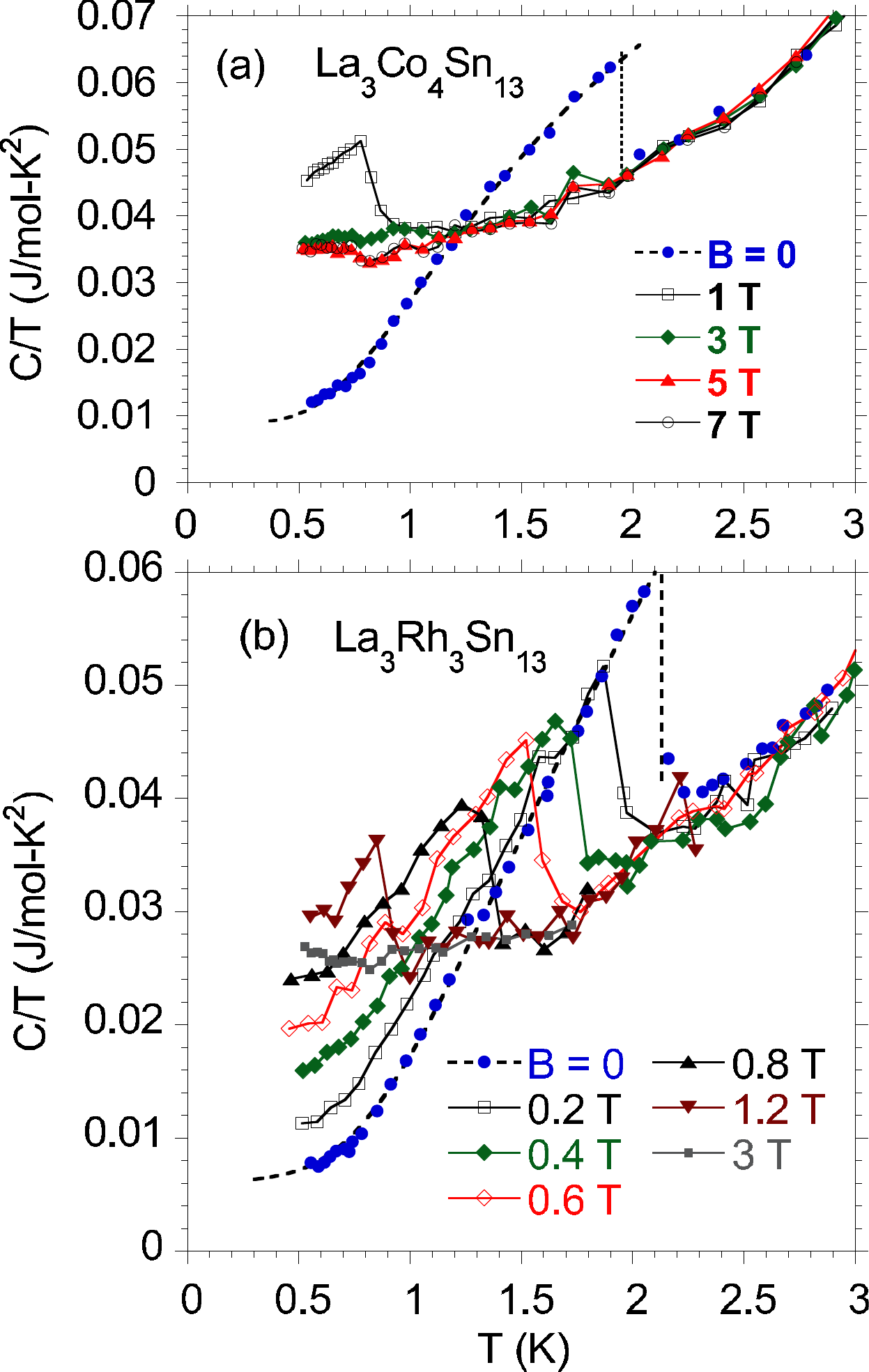}
\caption{\label{fig:Fig5}
(Color online) Temperature dependence of specific heat, $C(T)/T$, of La$_{3}$Co$_{4}$Sn$_{13}$ (a) and La$_{3}$Rh$_{4}$Sn$_{13}$ ($b$) at different magnetic fields.  The dotted line is the best fit of the expression $C(T)/T = \gamma + \beta T^2 + A\exp(-\Delta(0)/k_BT)$ to the data.
}
\end{figure}

\begin{figure}[h!]
\includegraphics[width=0.48\textwidth]{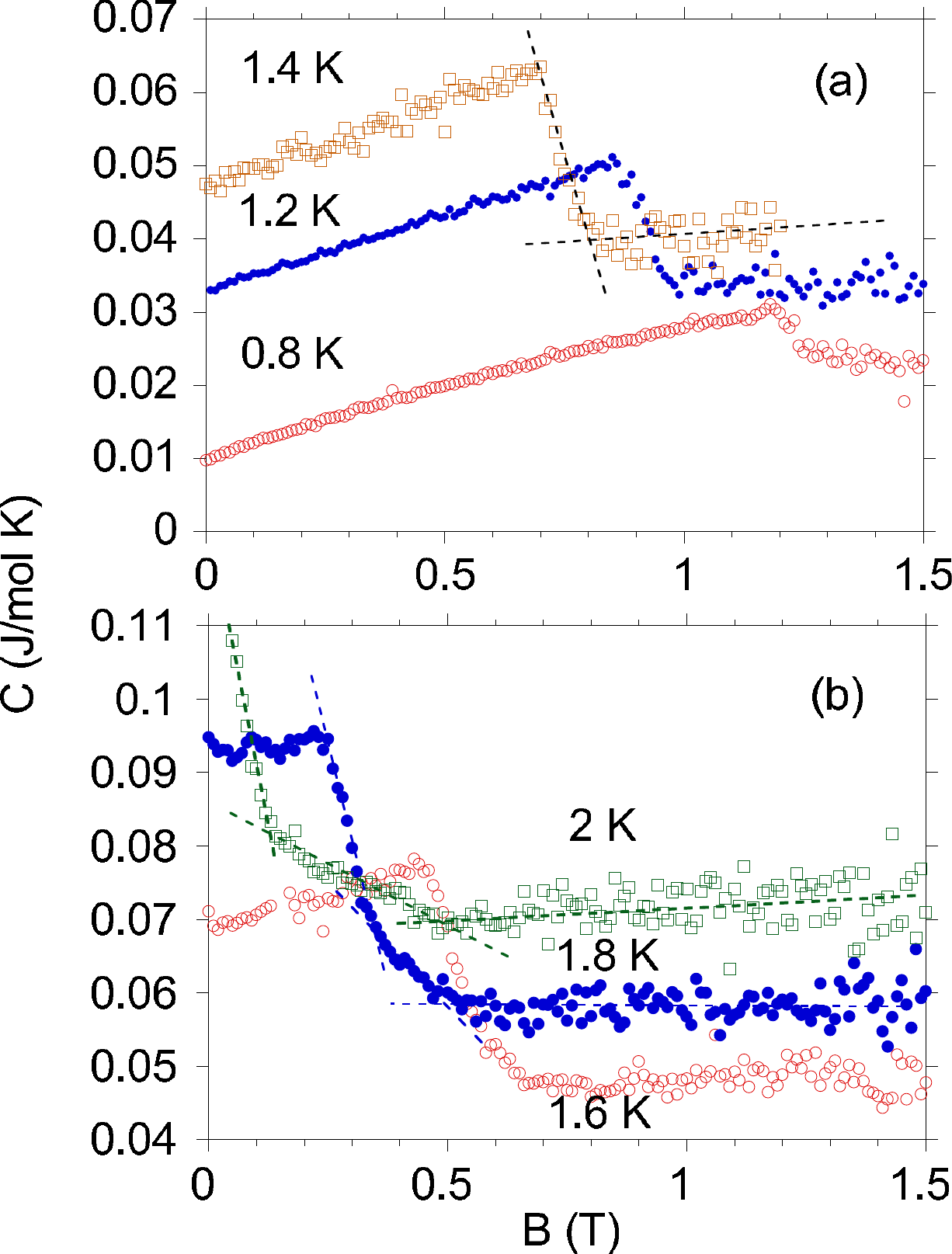}
\caption{\label{fig:Fig6} (Color online) Heat capacity vs magnetic field at constant temperature.}
\end{figure}

Shown in Fig.~\ref{fig:Fig5} is the specific heat $C$ of La$_{3}$Co$_{4}$Sn$_{13}$ and La$_{3}$Rh$_{4}$Sn$_{13}$ plotted as $C/T$ vs $T$ at various magnetic fields.
The heat capacity data indicate bulk superconductivity for La$_{3}$Co$_{4}$Sn$_{13}$ below $T_c = 1.95$ K [Fig.~\ref{fig:Fig5}(a)] in agreement with the electrical resistivity measurements, while the superconductivity in $C/T$ data for La$_{3}$Rh$_{4}$Sn$_{13}$ [in Fig.~\ref{fig:Fig5}($b$)] occurs below $T_c = 2.13$ K, in contrast to $T_c^{\star} = 2.98$ K obtained from electrical resistivity.  We note that in the $H-T$ phase diagram presented in Fig.~\ref{fig:Fig4}($b$), the $T_c$s significantly deviate from the GL theory in the temperature region $T \gtrsim 1.4$ K.  We therefore measured the heat capacity vs magnetic field at $T = {\rm const}$ for $T \leq 1.4$ K [Fig.~\ref{fig:Fig6}(a)] and $T >$~1.4 K [Fig. \ref{fig:Fig6}($b$)] to obtain the missing points in the $H-T$ phase diagram.
In  Fig.~\ref{fig:Fig6}(a), the heat capacity $C(H,T \leq$~1.4 K) has only one kink at $T_c$, while the $C$ data for $T > 1.4$ K in Fig.~\ref{fig:Fig6}($b$) show a kink at $T_c$ and at $T_c^{\star}$.
\begin{figure}[h!]
\includegraphics[width=0.48\textwidth]{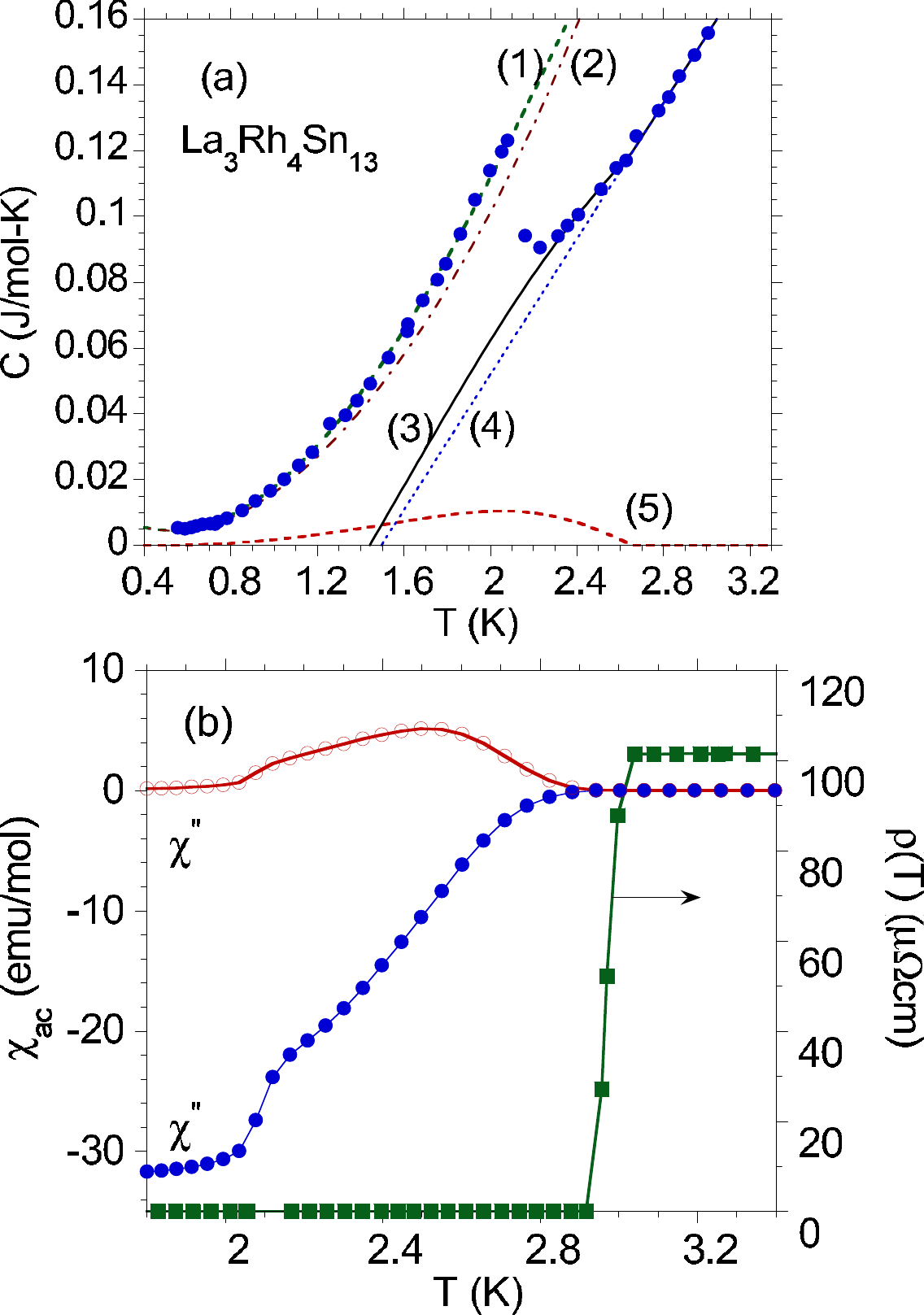}
\caption{\label{fig:Fig7} (Color online) Specific heat $C$ of La$_{3}$Rh$_{4}$Sn$_{13}$ approximated by the atomic-scale pair disorder model (a). Distinct specific heat contributions $C_i$ represent various combinations of lattice and electronic contributions as described in the text, where $i = 1-5$ and $C_4 + C_5 = C_3$ and $C_5 + C_2=C_1$.  For the purpose of comparison, the ac susceptibility and electrical resistivity are displayed in panel ($b$).
}
\end{figure}
These critical temperatures are both shown in the $H-T$ phase diagram [c.f. Fig.~\ref{fig:Fig4}($b$)] as blue filled ($T_c$) and unfilled ($T_c^{\star}$) squares, respectively.  Then, the $H$ vs $T_c$ dependence is well approximated by the GL theory with $H_{c2}(0) = 1.6$ T ($\xi_0=14$ nm).  The higher temperature superconducting phase between $T_c$ and $T_c^{\star}$ is interpreted in the context of electronic disorder over length scales similar to the coherence length, which is often observed in the high $T_c$ superconductors.

In Fig.~\ref{fig:Fig5}, the $C(T)/T$ data are fitted by the expression $C(T)/T = \gamma + \beta T^2 + A \exp(-\Delta (0)/k_BT)$.  The dotted curve represents the best fit with the fitting parameters obtained, respectively, for La$_{3}$Co$_{4}$Sn$_{13}$ and La$_{3}$Rh$_{4}$Sn$_{13}$: $\gamma$ = 9 mJ/molK$^2$ and 6 mJ/molK$^2$, $\beta$ = 2 mJ/molK$^4$ and 3 mJ/molK$^4$, and $\Delta(0) = 3.5$ K and 4.4 K, where $\Delta(0)$ is the energy gap at zero temperature.  From $\beta = N(12/5)\pi^4R\theta_D^{-3}$, we estimated the Debye temperature $\theta_D \sim 268$ K for La$_{3}$Co$_{4}$Sn$_{13}$ and 234 K for La$_{3}$Rh$_{4}$Sn$_{13}$.

Under zero magnetic field, $C(T)$ exhibits exponential $T$-behavior, which indicates $s$-wave superconductivity in both compounds.  BCS theory in the weak-coupling limit provides a relation between the jump of the specific heat at $T_c$ and the normal state electronic contribution, $\gamma$; i.e., $\Delta C/(\gamma T_c) = 1.43$.  We estimated $\Delta C/(\gamma T_c) \cong 1.5(5)$ for La$_{3}$Co$_{4}$Sn$_{13}$, taking the appropriate quantities as derived from Fig.~\ref{fig:Fig5} and the electronic specific heat coefficient $\gamma = 26$ mJ/molK$^2$ obtained below $T_c$ at the field 3 T.  In the case of La$_{3}$Rh$_{4}$Sn$_{13}$, $\Delta C/(\gamma T_c) \cong 1.5(5)$ using $\gamma = 15$ mJ/molK$^2$.  Moreover, the specific heat data give $2\Delta (0)/k_BT_c$ ratio values of 3.6 and 4.1 for Co and Rh samples, respectively, which are comparable with those expected from the BCS theory ($2\Delta(0)/k_BT_c = 3.52$).  We conclude that both compounds are typical BCS superconductors (c.
 f. Ref.~\onlinecite{Kase2011}) below $T_c$; however, in La$_{3}$Rh$_{4}$Sn$_{13}$, we found a second inhomogeneous superconducting phase between $T_c$ and $T_c^{\star}$ in magnetic fields lower than 1 T.  This phase is reduced by an applied pressure, as will be shown below.

\subsection{Possible explanation of the superconducting properties of La$_{3}$Rh$_{4}$Sn$_{13}$}

We clearly see two zero-field phase transitions in La$_{3}$Rh$_{4}$Sn$_{13}$ at temperatures $T_c = 2.13$ K and $T_c^{\star} = 2.98$ K.  Note that this observation strongly contrasts with the results on La$_{3}$Co$_{4}$Sn$_{13}$ where all the data (specific heat as well as electrical resistivity and magnetic susceptibility measurements) point to a well-defined single superconducting phase.  In the case of La$_{3}$Rh$_{4}$Sn$_{13}$, we observe a sharp jump of the specific heat at $T_c$ (see Fig.~\ref{fig:Fig5}), that indicates the sample is of good quality.  On the other hand, at the higher critical temperature $T_c^{\star}$, there is only a change of the slope of $C(T)$, as can be seen in Fig.~\ref{fig:Fig7}(a).  Therefore, magnetic susceptibility and electrical resistivity are shown in Fig.~\ref{fig:Fig7}(b) to help explain the physical properties of the the system below $T_c$ and  $T_c^{\star}$.  A sharp drop in the electrical resistivity is observed at $T_c^{\star}$.  
 The sharpness of this drop is very suggestive and it would be hard to imagine any other mechanism than superconductivity behind such a transition.  However, this sharp drop is not accompanied by any dramatic change of the magnetic susceptibility.  Instead, $\chi$ gradually decreases in the temperature window between $T_c$ and $T_c^{\star}$, and saturates first below $T_c$.  The saturation is accompanied by a specific-heat jump with a magnitude which remains in agreement with the BCS theory.  Note also that both transition temperatures decrease linearly with the application of external field (see Fig.~\ref{fig:Fig4}).  These linear dependencies, which hold in quite a broad range of temperatures, are hallmarks of the diamagnetic breaking of Cooper pairs, cf. the results of the Ginzburg-Landau theory or the Werthamer-Helfand-Hohenberg solution of the Gor'kov equations.

The presence of two sharp superconducting-like transitions can straightforwardly be explained in terms of two distinct superconducting phases, which are separated in space and/or involve different energy bands.  The onset of the low-temperature phase is accompanied by a jump in $C(T)$ and saturation of $\chi$.  The high-temperature phase occupies a much smaller volume of the system, but still allows for dissipationless charge currents, e.g., through a percolation-like transport.  The low-temperature phase is more robust against external magnetic fields, hence both transition temperatures become equal for $H \simeq 0.9$ T as shown in Fig.~\ref{fig:Fig4}.  For even stronger fields, the dominant phase completely masks the other phase.

It is quite clear that both phases have similar superconducting properties, i.e., comparable transition temperatures and upper critical fields.  One may speculate that these phases differ by the presence/magnitude of the distortion reported previously in Refs.~[\onlinecite{Slebarski13,Liu13}].  The distortion lowers the density of states at the Fermi level which decreases the transition temperature.  However, such a distortion simultaneously affects the Landau orbits, which effectively increases the upper critical field.\cite{MM-MM-2002}  While a microscopic theory is missing, one may carry out phenomenological modeling of $C(T)$ as discussed below.

The lack of a specific heat jump at $T_c^{\star}$ suggests that the high-temperature phase is spatially inhomogeneous.  A similar temperature dependence of the specific heat has been observed in PrOs$_4$Sb$_{12}$, where a double superconducting phase transition has been identified at temperatures $T_{c1} \approx$ 1.85 K and $T_{c2} \approx$ 1.70 K;\cite{Maple02,Vollmer03} in a few cases, a single sharp transition at $T_{c2}$ has been reported,\cite{Seyfarth06,Measson08} in a study in which Ru was partially substituted for Os, the transition at $T_{c1}$ was stabilized,\cite{Frederick07} and in other experiments only a broad peak in $C(T)$ was observed.\cite{Seyfarth06,Andraka10,Measson08}  The values of the lower and higher transition temperatures and the magnitudes of the corresponding specific heat jumps are sample dependent, suggesting sample inhomogeneity may be the origin of the double superconducting phase transition in PrOs$_4$Sb$_{12}$.

Another system where two superconducting phase transitions were found is CePt$_3$Si.  It is argued that the second transition results from a second phase with a slightly different chemical composition.\cite{Kim05}  It was reported that bulk superconductivity in a high-quality single crystal has a critical temperature significantly lower than for a polycrystal,\cite{Takeuchi07} which may suggest that the high-temperature superconductivity is related to disorder.  Moreover, the presence of inhomogeneities of superconducting characteristics has been reported even in a single crystal.\cite{Makuda09}  This may explain the fact that in high-quality single crystals, the electrical resistivity drops to zero at a temperature similar to the critical temperature of polycrystals.\cite{Takeuchi07}  This situation resembles what we observe in La$_{3}$Rh$_{4}$Sn$_{13}$, where the electrical resistivity drop is observed at a higher temperature than the temperature of the sharp jump in the sp
 ecific heat, indicating the onset of a bulk, homogeneous superconducting phase.

Anomalies similar to those observed in La$_{3}$Rh$_{4}$Sn$_{13}$ are present also in the specific heat of CeIrIn$_5$.  One of them, corresponding to the onset of bulk superconductivity, is a sharp jump in $C(T)$, whereas the other at higher temperature, is much less pronounced and corresponds to the drop of the electrical resistivity to zero.\cite{Bianchi01}  In this case, the discrepancy between $T_c$s determined from different measurements was also explained by the presence of an inhomogeneous superconducting phase.

Assuming that the scenario of inhomogeneous superconductivity in La$_{3}$Rh$_{4}$Sn$_{13}$ is plausible, we can explain the presence of the two anomalies in the specific heat at $T_c$ and $T_c^{\star}$.  Namely, we believe that the anomaly at $T_c^{\star}$ marks the onset of an inhomogeneous superconducting phase with spacial distribution of the magnitude of the superconducting energy gap $\Delta$. Following Ref.~\onlinecite{gabovich}, we assume a simple Gaussian gap distribution,
\begin{equation}
f(\Delta)\propto \exp\left[-\frac{\left(\Delta-\Delta_0\right)^2}{2d}\right],
\end{equation}
where $\Delta_0$ and $d$ are treated as fitting parameters.  The electronic contribution to the specific heat within the BCS theory can be given by the dashed line (5) in Fig.~\ref{fig:Fig7}(a).  The fitting parameters were determined in such a way that this electronic contribution, when added to the linear $C(T)$ observed above $T_c^{\star}$ [dashed line (4) in Fig.~\ref{fig:Fig7}(a)], describes the specific heat for temperatures between $T_c$ and $T_c^{\star}$ [line (3)].  Of course, the inhomogeneous phase contributes also to the specific heat below $T_c$.  This means that the experimental data for the specific heat below $T_c$, fitted by the dashed line (1), includes both the contributions from the homogeneous and inhomogeneous phases.  Subtracting the inhomogeneous contribution, we obtain the dash-dotted line (2), which represents only the homogeneous phase with a spatially uniform energy gap.  The absence of a significant anomaly in the specific heat at $T_c^{\star}$ suggests that only a small part of the sample becomes superconducting at the higher critical temperature.  This volume fraction, however, has to be large enough to produce a complete drop of the electrical resistivity to zero.

The assumption of the existence of homogeneous and inhomogeneous regions with different critical temperatures
seems to explain both the double superconducting phase transition and the
shape of the specific heat. The question however remains, what induces
the inhomogeneity of the superconducting order parameter? One of the possible
explanations is the presence of a small number of impurities. This scenario
is supported by the fact that the critical temperature in the inhomogeneous
phase is {\em higher} than in the homogeneous one. It has been shown in Ref. 
\onlinecite{inhomo}
that in strongly correlated systems the magnitude of the superconducting
order parameter can be {\em increased} in a vicinity of an impurity. This mechanism
has been used to explain the observed enhancement of superconductivity close to off--plane
oxygen dopants in the high--$T_c$
superconductors\cite{McElroy} and may lead to a specific heat similar to that observed in 
La$_3$Rh$_4$Sn$_{13}$ between $T_c$ and $T^\star_c$.\cite{Andersen06}

In the above we assumed a scenario based on a presence of regions with inhomogeneous superconducting order parameter. One can also imagine other explanations, e.g., based on an assumption of a presence of completely different phase. However, since the resistivity sharply drops to zero in the high--critical--temperature phase the volume occupied by this phase would have to be large enough to be beyond the percolation threshold. But such a large volume of
a different phase would be clearly visible in ..... , what is not the case. Therefore, we exclude such a possibility.

\subsection{Electrical resistivity of La$_{3}$Rh$_{4}$Sn$_{13}$ and La$_{3}$Co$_{4}$Sn$_{13}$ under applied pressure}

Intermetallic superconductors often exhibit structural instabilities.\cite{Chu74}  The application of external pressure to these superconducting materials can drive the compounds towards (or away) from lattice instabilities by varying the dominant parameters determining the superconducting properties; e.g., electronic density of states at the Fermi level.  Generally, pressure is an important parameter as it can be used to analyze the usefulness of theoretical models (e.g., the case of the quantum critical behaviors observed at the quantum critical point in several heavy fermions with unconventional superconductivity, where pressure is a possible tuning parameter).  Most superconducting metals show a decrease of $T_c$ with pressure.\cite{Olsen64}  The pressure dependence of $T_c$ can be understood within the weak-coupling BCS model\cite{Bardeen57} or the Eliashberg theory of strong-coupling superconductivity.\cite{Eliashberg61}  The La-based compounds studied here are classifi
 ed as weakly coupled BCS-like superconductors.

\begin{figure}[h!]
\includegraphics[width=0.48\textwidth]{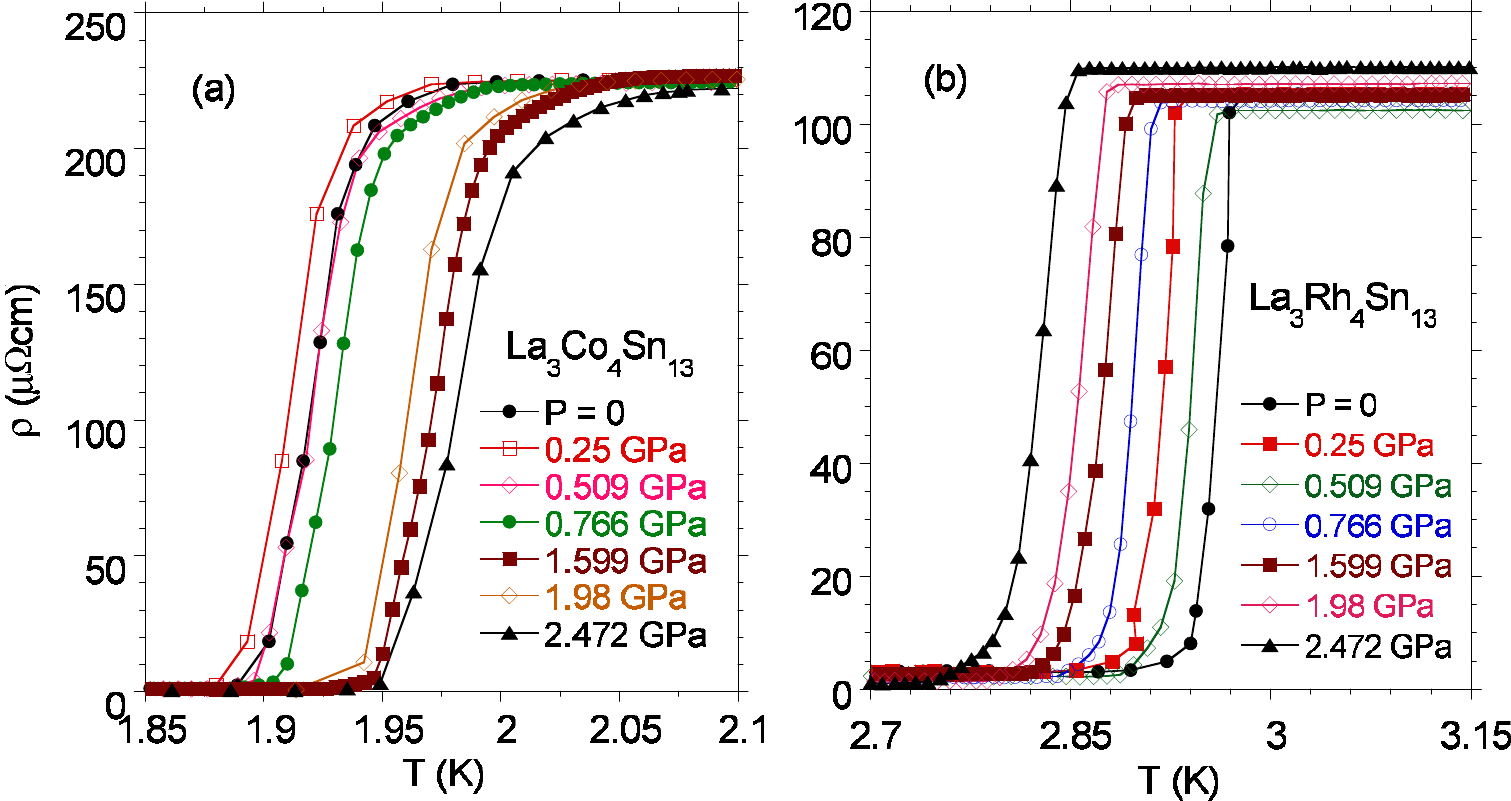}
\caption{\label{fig:Fig8} (Color online) Electrical resistivity of La$_{3}$Co$_{4}$Sn$_{13}$ (a) and La$_{3}$Rh$_{4}$Sn$_{13}$ (b) at different applied pressures.
}
\end{figure}

\begin{figure}[h!]
\includegraphics[width=0.48\textwidth]{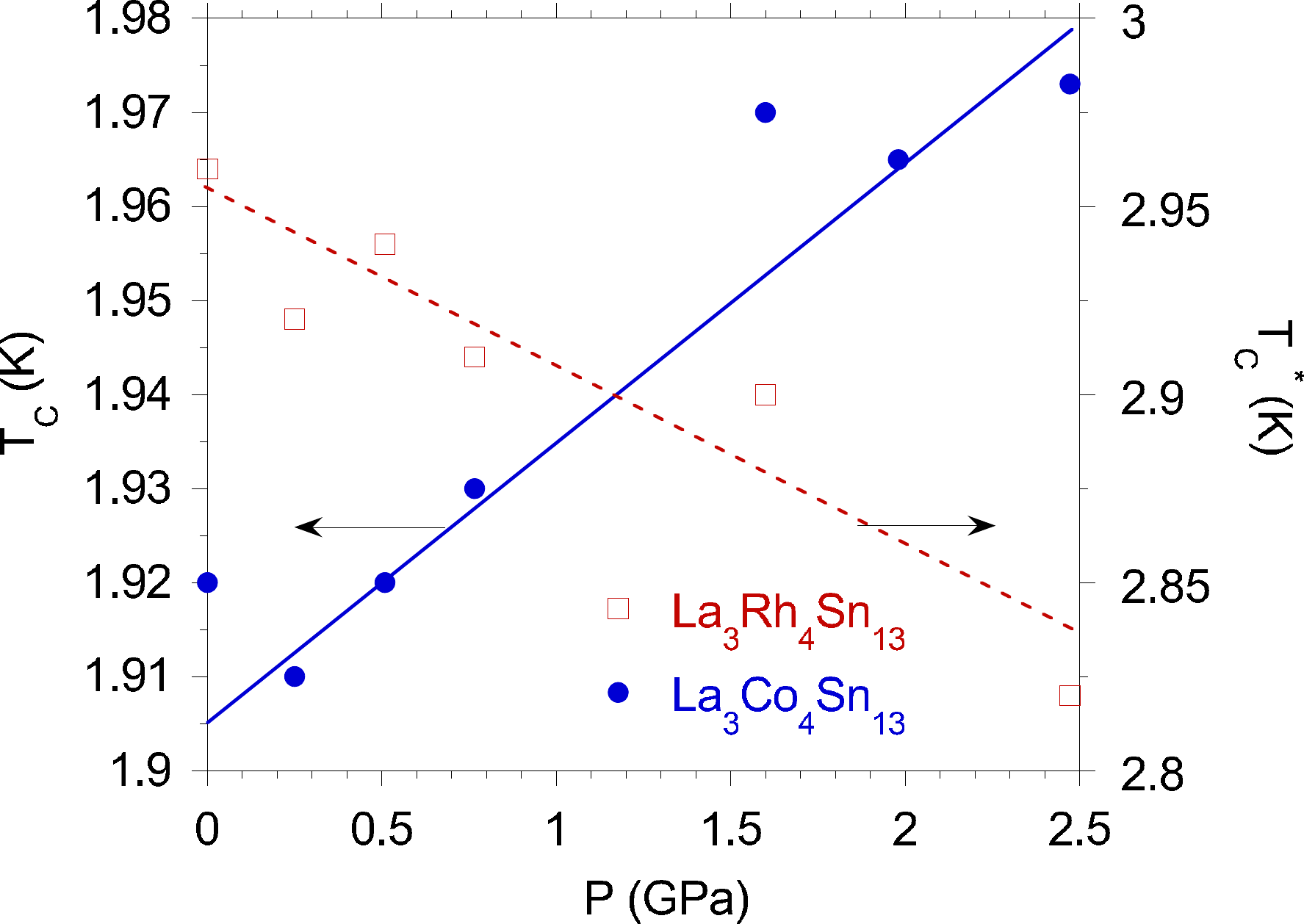}
\caption{\label{fig:Fig9} (Color online) Critical temperatures $T_c$ and $T_c^{\star}$ vs pressure $P$.  $T_c$s are obtained from electrical resistivity under $P$ and defined as the temperatures where $\rho$ decreases to 50\% of its normal-state value.
}
\end{figure}

\begin{figure}[h!]
\includegraphics[width=0.48\textwidth]{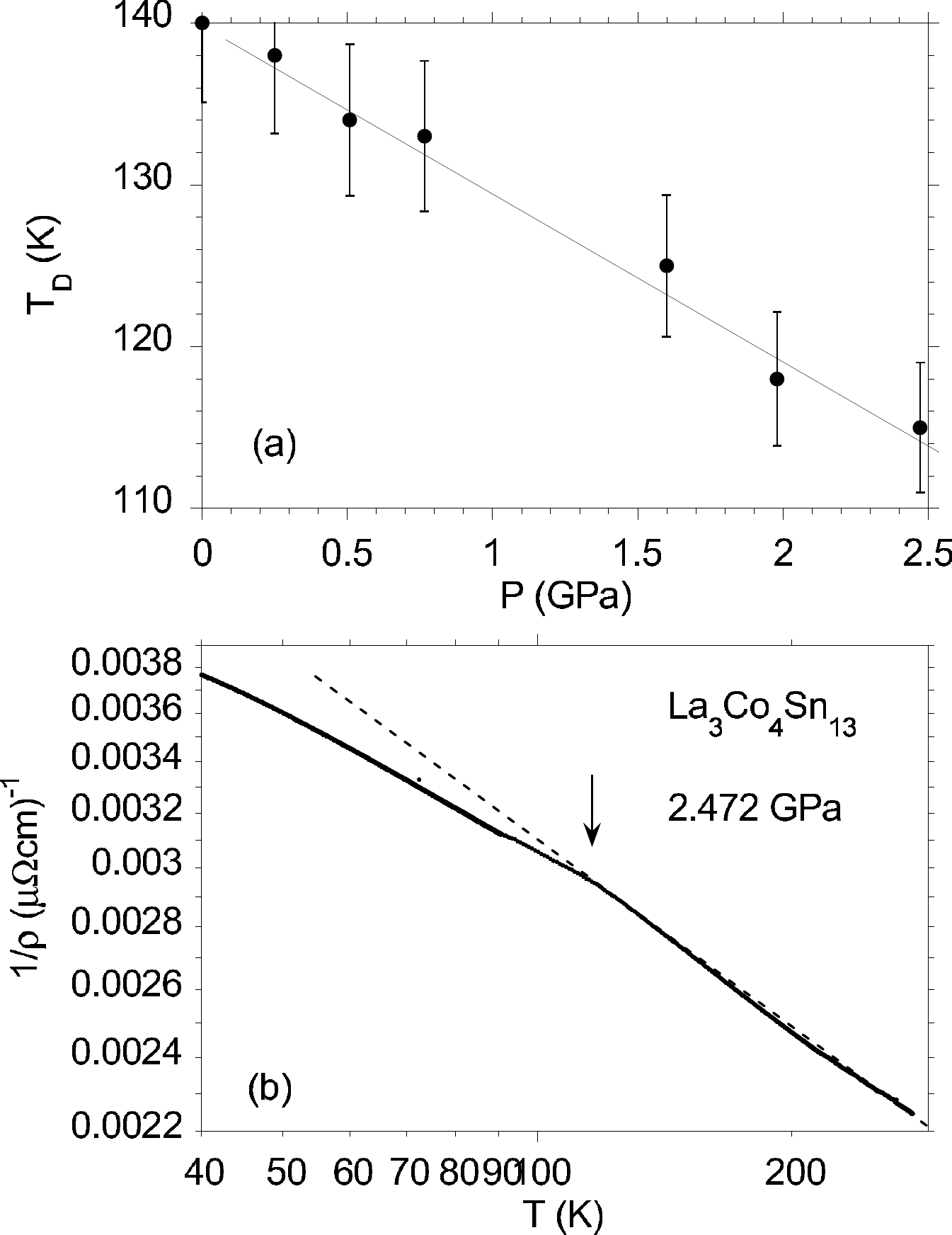}
\caption{\label{fig:Fig10} (a) Temperatures $T_D$ associated with a weak structural distortion in La$_{3}$Co$_{4}$Sn$_{13}$ vs pressure $P$.  $T_D$ was defined by a change of the slope of $1/\rho$ vs $T$ in a log-log scale [example is shown in panel (b) where an arrow emphasizes the slope change].
}
\end{figure}

Figures~\ref{fig:Fig8} and \ref{fig:Fig9} show the effect of applied pressure on the resistive critical temperatures for La$_{3}$Co$_{4}$Sn$_{13}$ and La$_{3}$Rh$_{4}$Sn$_{13}$.
The superconducting $T_c^{\star}$ of La$_{3}$Rh$_{4}$Sn$_{13}$ decreases linearly with applied pressure at a rate of $dT_c^{\star}/dP = -0.05$ K/GPa.  However, the $T_c$ values of La$_{3}$Co$_{4}$Sn$_{13}$ increase with pressure with a pressure coefficient of $\sim 0.03$ K/GPa. 
One should take into account, however, that the transition temperature obtained from the resistivity characterizes two different superconducting phases, namely a homogeneous phase for La$_{3}$Co$_{4}$Sn$_{13}$ and an inhomogeneous one for La$_{3}$Rh$_{4}$Sn$_{13}$, which means that $T_c^{\star}$ vs $P$ is not an intrinsic behavior in the high pressure phase diagram. Moreover, in the high pressure phase diagram $T_c^{\star}$ decreases only slightly with increasing of applied pressure, staying well above the critical temperature $T_c$ found from the specific heat. Therefore in the following we discuss the $T_c$ vs $P$ behavior and its consequences on the physical properties of La$_{3}$Co$_{4}$Sn$_{13}$.
While the superconductivity in both compounds is well described by the BCS theory in the weak-coupling limit, this unusual positive $dT_c/dP$ behavior has been observed in superconducting materials with lattice instabilities; e.g., V$_3$Si,\cite{Chu74} which is a weakly coupled BCS-like superconductor and undergoes a small structural phase transition at $T_L > T_c$ from a high-temperature cubic phase to a low-temperature tetragonal phase.  It was proposed that soft phonon modes play a major role in stabilizing superconductivity.  To investigate an interplay between $T_c$ and soft phonon modes leading to structural instabilities, it is desirable to tune the temperature of the structural distortion to $
 T_c$ by chemical or applied pressure.  Detailed investigations (x-ray diffraction, resistivity vs temperature, etc.)\cite{Slebarski13,Liu13} indicated that a subtle structural distortion in La$_{3}$Co$_{4}$Sn$_{13}$ occurs at $T_D \sim 140$ K, characterized by a deformation of the Sn$_{12}$ cages and accompanied by Fermi-surface reconstruction.  In Fig.~\ref{fig:Fig10}(b), each value of $T_D$ was defined as the temperature where the resistivity $\rho$, plotted as $1/\rho$ vs $T$ in log-log scale, shows a kink.
In Fig.~\ref{fig:Fig10}(a), the system is superconducting for applied pressures $P < 2.5$ GPa.  We speculate that a superlattice quantum critical point could be observed in La$_{3}$Co$_{4}$Sn$_{13}$ at an applied pressure of $\sim 20$ GPa.  Such a scenario was reported for the related superconducting compound Ca$_3$Ir$_4$Sn$_{13}$ near a critical pressure of about 1.8 GPa.\cite{Klintberg12}

\section{Conclusions}

The compounds La$_{3}$Co$_{4}$Sn$_{13}$ and La$_{3}$Rh$_{4}$Sn$_{13}$ are BCS superconductors, which have been the subject of recent interest.\cite{Kase2011}  We have concentrated on studying the superconducting states of these compounds under applied pressure and magnetic field.  The first significant observation we have made is that La$_{3}$Rh$_{4}$Sn$_{13}$ exhibits two superconducting transitions in a weak magnetic field, characterized by two step--like drops in the ac magnetic susceptibility as a function of temperature below $T_c^{\star}$, by a sharp drop in the electrical resistivity at $T_c^{\star} = 2.98$ K, and by a significant discontinuity in the specific heat at $T_c = 2.13$ K ($T_c < T_c^{\star}$).  This complicated anomaly is interpreted in the context of the presence of an inhomogeneous superconducting phase between $T_c$ and $T_c^{\star}$.  Similar behavior has been observed for a few other superconducting heavy fermion systems including PrOs$_4$Sb$_{12}$, CeP
 t$_3$Si, and CeIrIn$_5$.  Our results contribute towards developing a broader understanding of this complex behavior in novel superconducting materials.  Second, we found an unusual pressure effect on $T_c$ in La$_{3}$Co$_{4}$Sn$_{13}$.  A positive $dT_c/dP$ behavior is discussed in the context of the presence of a structural instability near $T_D \sim$~140 K at ambient pressure, which strongly decreases with applied pressure.  A similar pressure-dependent structural change was observed in the isostructural and nonmagnetic compound Ca$_3$Ir$_4$Sn$_{13}$.  In this compound, $T_D$ is suppressed with applied pressure such that $T_D \rightarrow 0$ K near 2 GPa and a novel superlattice quantum critical point is observed.  We speculate that a similar quantum criticality for La$_{3}$Co$_{4}$Sn$_{13}$ could be observed at a pressure about one order of magnitude larger.

\section{acknowledgments}

One of us (A.\'S.) is grateful for the hospitality at the University of California, San Diego (UCSD).  He also thanks the National Science Centre (NCN) for financial support, on the basis of Decision No. DEC-2012/07/B/ST3/03027.  One of us (M.F.) thanks the National Science Centre (NCN) for financial support, on the basis of Decision No. DEC-2011/01/N/ST3/03476. M.M.M. acknowledges support from the Foundation for Polish Science under the ``TEAM'' program for the years 2011–2014. Measurements under applied pressure and performed at UCSD were supported by the National Nuclear Security Alliance program through U.S. Department of Energy Grant No. DE-NA0001841.

\end{document}